\documentclass[final,5p,times,twocolumn]{elsarticle}

\usepackage{amssymb}

\usepackage{siunitx}

\usepackage[version=4]{mhchem}

\usepackage{booktabs}

\usepackage[caption=false,font=footnotesize]{subfig}

\usepackage{cleveref}

\journal{ } %
\newcommand{\cevns}{CE$\nu$NS}

\begin{document}

\begin{frontmatter}



\title{Submarine Navigation using Neutrinos}
\date{}

\author[inst1]{Javier Fidalgo Prieto}
\author[inst1]{Stefano Melis}
\author[inst1]{Ana Cezon}
\author[inst1]{Miguel Azaola}
\author[inst1]{Francisco José Mata}
\author[inst1]{Claudia Prajanu}

\affiliation[inst1]{organization={GMV},
            addressline={Isaac Newton 11 P.T.M. Tres Cantos}, 
            postcode={28760},   
            city={Madrid},
            country={España}}

\author[inst2,inst3]{Costas Andreopoulos}
\author[inst2]{Christopher Barry}
\author[inst2]{Marco Roda}
\author[inst2]{Julia Tena Vidal}

\affiliation[inst2]{organization={University of Liverpool, Department of Physics},
            city={Liverpool},
            postcode={L69 7ZE}, 
            country={UK}}
            
\affiliation[inst3]{organization={Science and Technology Facilities Council},
            addressline={Rutherford Appleton Laboratory, Particle Physics Dept.,}, 
            city={Oxfordshire},
            postcode={OX11 0QX}, 
            country={UK}}
            
\author[inst4]{Florin-Catalin Grec}
\author[inst5]{Luis Mendes}

\affiliation[inst4]{organization={ESA, ESTEC},
            addressline={Keplerlaan 1, PO Box 299}, 
            postcode={NL-2200 AG Noordwijk}, 
            country={The Netherlands}}            
            
 \affiliation[inst5]{organization={Rhea System SA for ESA},
            addressline={ESAC}, 
            city={Camino Bajo del Castillo s/n},
            postcode={Urb. Villafranca del Castillo }, 
            state={28692 Villanueva de la Canada (Madrid)},
            country={SPAIN}}           

\begin{abstract}
Neutrinos are among the most abundant particles in the universe, nearly massless, travel at speeds near the speed of light and are electrically neutral. Neutrinos can be generated through man-made sources like particle accelerators or by natural sources like the sun. Neutrinos only interact via the weak force and gravity. Since gravitational interaction is extremely weak and the weak force has a very short range, neutrinos can travel long distances unimpeded through matter, reaching places inaccessible to GNSS (Global Navigation Satellite System) signals such as underwater locations. The main objective of this work is to sketch an early high-level design of a Neutrino PNT (Position, Navigation and Timing) mission and analyze its feasibility for submarine navigation since there is a need to improve current navigation technologies for submarines. The high-level preliminary concept proposes Cyclotrons or Linear Accelerators based on the physical process Pion Decay at Rest as neutrino sources. For detecting such isotropic neutrino fluxes user equipment must be composed of a high-performance clock synchronized with the system, a detector and possibly additional sensors such as IMU (Inertial Measurement Unit). A feasibility analysis of the recommended system option is performed based on simulations for determining the neutrino detection rate and on a PNT tool to estimate the PNT performances. Although the submarine navigation application is in the limit of being feasible with current technology, it could be realized with some important but reasonable progress in source and neutrino detector technology.
\end{abstract}




\end{frontmatter}


\section{Introduction}
\label{sec:intro}
GNSS (Global Navigation Satellite System) has a large number of applications and is an indispensable utility in multiple diverse fields, ranging from navigation to surveying, data observation, timing and reconnaissance. It is sometimes called the invisible utility and has both civilian and military functions. As more and more applications were being developed and users became more dependent on the service, some of the weaknesses of satellite navigation came to light. For example, the inherent characteristics of the signal meant that in indoors, urban environments, underwater, or underground it is not possible to determine position of a user. In addition, GNSS signals are susceptible to jamming or spoofing. Although urban and indoor positioning can be solved with additional sensors and positioning systems such as 5G, WiFi, etc., an investment in new infrastructure may be required to enable positioning services underwater and underground.

Using neutrino sources as navigation beacons could enable navigation and tracking directly through normal matter. This property is capturing interest for applications in environments where, today, no other suitable navigation means is available like in the case of underwater and underground. This project explores under which circumstances a neutrino-based PNT concept can guide submarines and describes what a neutrino PNT system may entail.

\section{Motivation}
Neutrinos possess two properties that make them very attractive for communications and navigation in subsurface or underwater environments: they interact very weakly with matter and travel with velocities very close to the speed of light.
 The former means that neutrinos are very difficult to detect implying the need of very large and heavy detectors, which represent a challenge for potential applications when considered from an operational standpoint. Miniaturization of neutrino detectors is then a key goal for the practical application of neutrinos.
This paper focus on Submarine Navigation but other interesting use cases for Neutrino PNT concept can be identified, such as subsurface PNT, back-up of GNSS in denied environments or spacecraft positioning or tracking.

\subsection{Previous Works} \label{sec:previous_works}
The use of neutrino beams for communication is an old idea and has been put forth by several authors and for various purposes. Different activities have been undertaken in the past by different research organizations, including NASA (National Aeronautics and Space Administration) and Universities, with the goal of using neutrinos for communication or PNT applications.

Demonstration of communication based on neutrinos has been executed at FERMILAB \cite{Stancil}. This experiment was performed using the Neutrino beam at the main Injector (NuMI) as neutrino source (data sender) and the MINER$\nu$A detector as receiver. The distance between the first stage of the NuMI and the MINER$\nu$A detector is 1035 m. The communication link achieved a decoded data rate of 0.1 bits/s with a bit error rate of 1\%.

In \cite{Huber} it was proposed to use neutrinos from the decays of muons stored in a muon storage ring to establish communication with submarines. In order to achieve a sufficiently large cross section the storage muon energies should be about 150 GeV. Assuming a rate of $10^{14}$ $s^{-1}$ this implies a 4 MW proton beam, 2.4 MW for acceleration which translates into an electrical power consumption of 65 MW (the energy produced by a small power plant).

Ideas to exploit solar neutrino particles for navigation have been even patented \cite{Gutt}. This patent describes a method to navigate using celestial/solar neutrinos based on the neutrino arrival angle.
The main obstacle to use neutrinos for submarine navigation is that the low detection rate which could force to have rather big detectors which may not fit in a submarine nor in size neither in weight.

However, the recent experimental confirmation \cite{Akimov2017} of the  coherent elastic neutrino-nucleus scattering ({\cevns}), predicted for the first time more than 40 years ago, has opened a new perspective \cite{Akimov2017}. {\cevns} is realized when a neutrino interacts with a nucleus as a whole, implying very small nucleus recoil momentum (in the keV range) and neutrino energy below 50 MeV. The consequence is a rather big cross-section compared to non-coherent scattering at the same energy. This allows the construction of smaller detectors with smaller detection times. For instance the COHERENT collaboration performed its study using a detector consisting of 14.6 Kg sodium sopped CsI scintillator. This is a very important technological evolution in the context of neutrino PNT since the main obstacle is the size of the detectors which usually are not portable because of their large size, besides the small detection rate. 

\subsection{Submarine Navigation}

The use of neutrinos for communication with a submarine deep under the ocean has been previously considered \cite{Huber}. In specific environments, for example those where nuclear powered submarines operate, there is a requirement for unlimited submerged endurance. However, the need to position themselves and also to communicate may force them ultimately to come to the surface at some point, as currently there are still no technical solutions to ensure  communication at operational speeds and for the depths of operations. In \cite{Huber} it has already been discussed how a neutrino beam from a muon storage ring could be detected by sensors mounted on the hull of a submarine, allowing the establishment of a one-way communication link at speed and depth with data rates in the range from 1 to 100 bits per second. This could improve current communication technologies based on Extremely Low Frequencies (ELF) and Very Low Frequencies (VLF), by a factor of 1 to 3 orders of magnitude.

Other innovative navigation concepts for submarines are being explored in the literature, such as gravity gradient terrain matching to improve navigation with submarines.

For navigation using neutrinos, the Submarine would have to carry a Neutrino detector. Neutrino detector weight has been considered in this work according to the operational restrictions of the application, but on the other hand, large enough to enhance the number of detected neutrinos. 
Typical submarine weights are of the order 10.000 tones and then, neutrino detector weight could only be a small fraction of the total submarine weight, taking also into account that the need for detector shielding could increase the total weight of the infrastructure. When developing the Neutrino PNT concept it has also been considered that this navigation method would be supplemented with other navigation sensors currently in use such as Inertial Measurement Unit (IMU). The performance of IMU degrades in time and the Neutrino PNT concept would allow to periodically re-establish - each time a new neutrino is detected - the IMU error within the accuracy requirement. Realistic values for the IMU performance have been  considered in the simulations for the proof-of-concept of the Neutrino PNT design. For performance assessment a total positioning error no larger than 1 km has been considered as requirement associated to submarine navigation, assuming that collision avoidance is achieved by other sensors.

\section{Concept and System Architecture}

In order to fully characterize a neutrino based PNT system three items are of interest: means to generate neutrino particles, means to detect them, and a suitable positioning algorithm. After presenting a trade-off of technologies available to build such a system (Sec. \ref{sec:TradeOff}), we describe  a possible system architecture (Sec. \ref{sec:SystArch}), the user equipment (Sec. \ref{sec:UserEquip}) and a suitable positioning algorithm (Sec. \ref{sec:PNTAlg}).   
\subsection{Architecture Trade-off}\label{sec:TradeOff}
In this section we will analyze the possible choices for the three pillars (sources, detectors, PNT algorithms) that can enable submarine navigation with neutrinos.   

Concerning the positioning algorithms, they can be divided in two broad categories based on their working principle: Time of Arrival (ToA), i.e. ranging techniques, or Angle of Arrival (AoA). Ranging algorithms need time synchronization of sources, and between users and sources in order to measure the transmission and reception time in the same time reference. AoA algorithms instead need the reconstruction of the neutrinos' trajectories in order to localize the source or, inversely, the detector.
Apart from tagging the time or the angle of arrival, one can also consider encoding information (for instance, time and source position) in the ``signal" following GNSS working principle. In the case of a neutrino, information can be encoded in the form of bits where each bit is the presence/absence of a neutrino packet as done in \cite{Stancil}. 

Among available neutrino sources we find solar neutrinos (natural source), accelerators, and fission reactors (artificial sources). Accelerator sources can be further divided in two categories depending if the neutrino are produced through decay on the fly or decay at rest of the parent particle (typically a pion). In the first category enter all the accelerators able to produce a focused (non-isotropic) neutrino beam. This kind of sources can produce neutrinos at whatever energy is technologically possible and with fluxes potentially high. An example of such kind of accelerators is the already cited NuMI at FERMILAB~\cite{Stancil}. On the contrary "pion-decay at rest" based accelerators are characterized by a limited spectrum in energy and by an isotropic emission of the neutrinos. The flux of the neutrinos produced depends on the flux of the parent particles (mainly pions) and on the capability to stop the parent particles on the target which {\it de facto} limits the design, the energy and the flux. Neutrino production through accelerators can be modulated, this means that it is technically possible to embed information in the generated neutrinos as commented before.

Fission nuclear reactors can produce abundantly anti-neutrinos.  Their energy spectrum is determined by the fuel used. Their flux is isotropic and limited by the quantity of fuel/number of reactors. In this case it is rather difficult to embed information in the produced anti-neutrino flux because the only way to modulate the production is shutting down the reactor which is a complex operation with a large latency.
Notice that the isotropic properties of ``pion decay at rest" produced neutrino or of reactor anti-neutrinos is a strong plus for a PNT system because this ensures the possibility to have a global coverage while in the case of focused beam only a small volume would be "illuminated".

In the case of natural sources like solar neutrinos it is evident from the beginning that it is not possible to embed any information in the signal nor using some ranging technique, due to the absence of proper time synchronization. AoA algorithms would be the only possibility. However, because of the energies that dominate the solar neutrino spectrum it is difficult to reconstruct trajectories in present detectors. Therefore this kind of source has been discarded.
 
Neutrino detectors for scientific applications are in general very large (the size of a building), see for instance Super-Kamiokande \cite{Fukuda:2002uc} or sometimes huge as \SI{1}{\kilo\metre\cubed} like the IceCube project~\cite{Aartsen:2016nxy}, heavy (Super-Kamiokande is about 50000 tons) and very expensive. Smaller detectors are used only in the very proximity of the source, like anti-neutrinos detectors for nuclear reactors, see for instance ~\cite{Carroll_2015}. Miniaturization of detectors is necessary to ensure mobility and portability together with cost reductions. Presently, the {\cevns} based detectors are very promising thanks to the {\cevns} enhanced interaction cross-section which scales as the square of the number of nucleons in the nucleus, thus increasing the cross section up to 2 orders of magnitude in typical crystal based detectors~\cite{Akimov2017}, see also Fig.~\ref{fig:CEvNS_xsec}.
 
From the description above, the best solution  consists in accelerators where neutrinos are produced through pion decay at rest coupled to {\cevns} based detectors (which overlap perfectly in the energy range) using a ranging algorithm. Notice that also reactor anti-neutrinos can be detected using {\cevns} technique. However, the main challenge is related with the flux that may not be sufficient to detect neutrinos from a submarine 1-1000 Km apart and may not be incremented significantly. Similarly, beam based accelerators have been discarded since they have a limited coverage.

The solution of this high-level trade-off has been exploited to build our system architecture. 

\subsection{Concept and System Architecture}\label{sec:SystArch}
\begin{figure}[!t]
\centering
\includegraphics[width=2.5in]{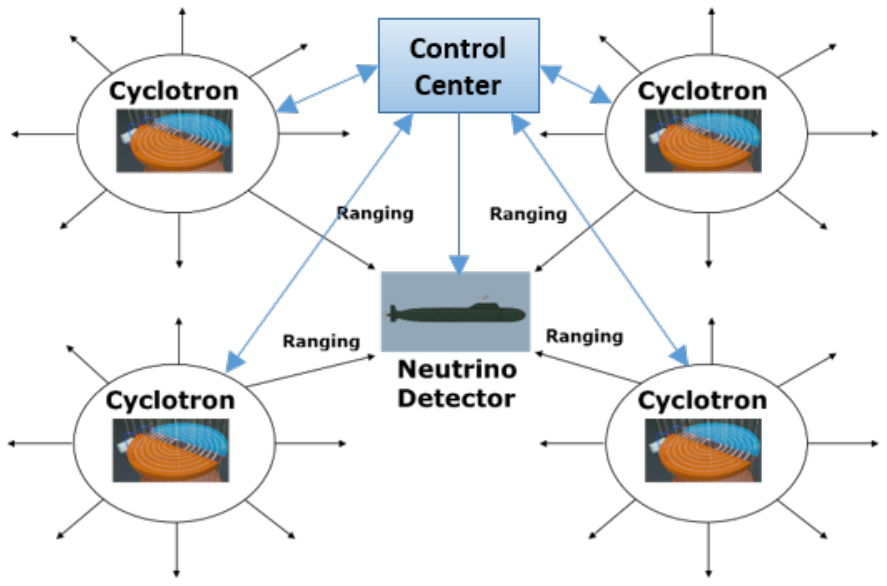}
\caption{Concept and system architecture.}
\label{fig:concept_system_arc}
\end{figure}
Our neutrino PNT system architecture is depicted in Fig.~\ref{fig:concept_system_arc}. Our design provides for an infrastructure segment, formed by a lattice of neutrino sources and a control center, and a user segment, formed by neutrino detectors equipped with (atomic) clocks and IMU. 
As commented before, neutrino sources are accelerators that produce neutrinos through pion decay at rest. An example of such kind of accelerators is that proposed for the \mbox{DAE$\delta$ALUS} (Decay at Rest Experiment for \mbox{$\delta$cp} studies At the Laboratory for Underground Science) experiment ~\cite{Conrad2009}. \mbox{DAE$\delta$ALUS} accelerator is based on (two steps) cyclotrons acceleration with an innovative design that allows  a neutrino flux of \num{4e22}~\(\nu\)/year/flavour, see Sec.~\ref{sec:neutrino_sources}. Cyclotrons are very compact accelerators (few meters of diameter). They represent a very good choice to deploy a large number of generators or even re-deploying an existing one to a new location.
The control center will coordinate the various sources in order to maintain the synchronization of the generator, monitoring the position of the neutrino sources, disseminating notification to users etc..
The main functions of the user equipment would be, as usual, to detect neutrinos and compute the PNT solution as described in the next section.

\subsection{User Equipment}\label{sec:UserEquip}
\begin{figure}[!t]
\centering
\includegraphics[width=2.5in]{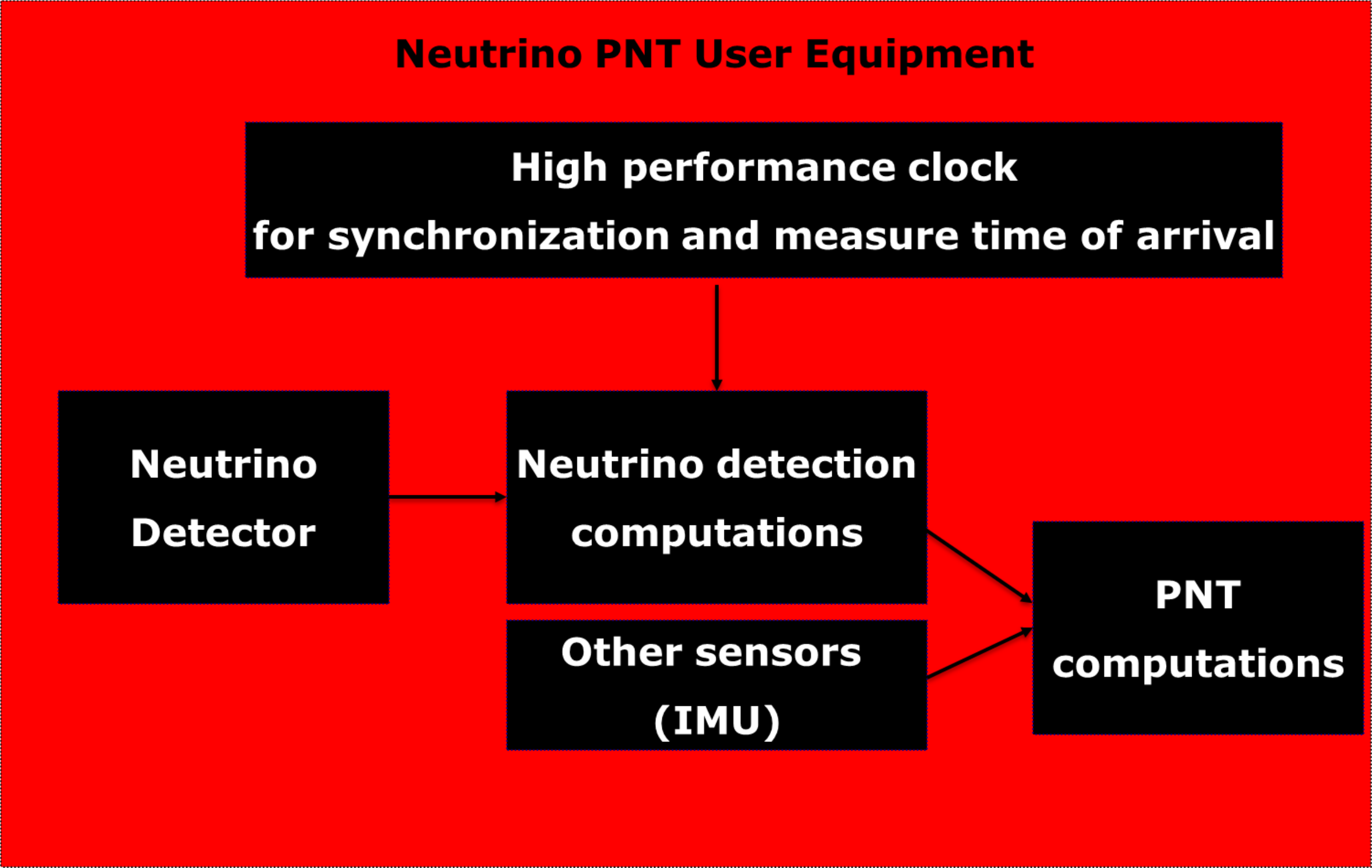}
\caption{User equipment.}
\label{fig:user_equip}
\end{figure}
User equipment design is depicted in Fig.~\ref{fig:user_equip}. It comprises:
\begin{itemize}
\item a neutrino detector suitable to detect the {\cevns} process like CsI[Na] crystal scintillators and all the ancillary systems (shielding, power supply etc.);
\item a computational unit performing  the data analysis after detection;
\item atomic clocks or any other high performance clocks to maintain detector clock synchronization to system time (needed to measure the neutrino time of arrival in the same time scale of the system);
\item any other sensor assisting navigation like IMUs;
\item a computational unit to perform PNT from neutrinos and other sensors data.
\end{itemize}
Notice that the presence of sensors like IMU is crucial, since from our simulations, the number of detected neutrinos in a typical scenario is not sufficient to make a trilateration in acceptable time window. More details in the next section and in Sec.~\ref{sec:sim_feasability}.

\subsection{PNT Algorithms}\label{sec:PNTAlg}
\begin{figure*}[!t]
\centering
\subfloat[]{\includegraphics[width=2.5in]{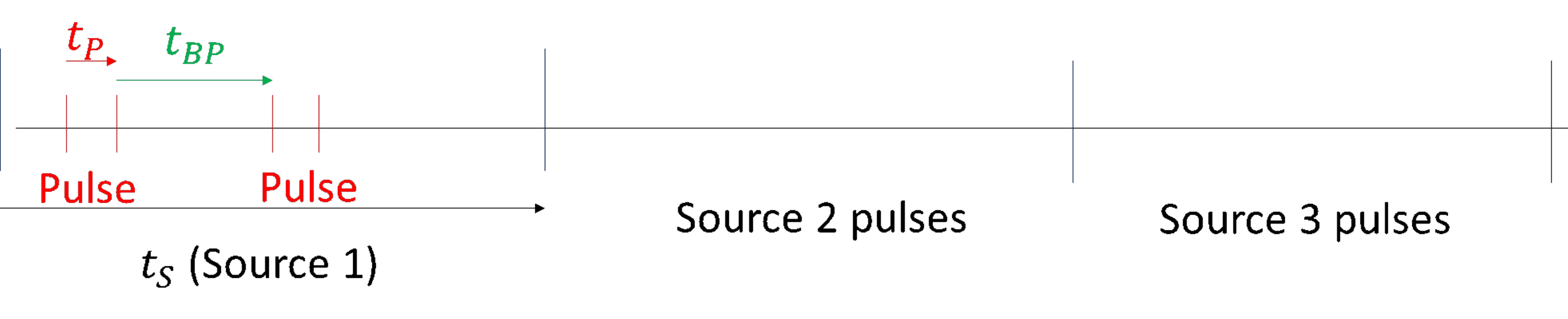}%
\label{fig:seq_trilat}}
\hfil
\subfloat[]{\includegraphics[width=2.5in]{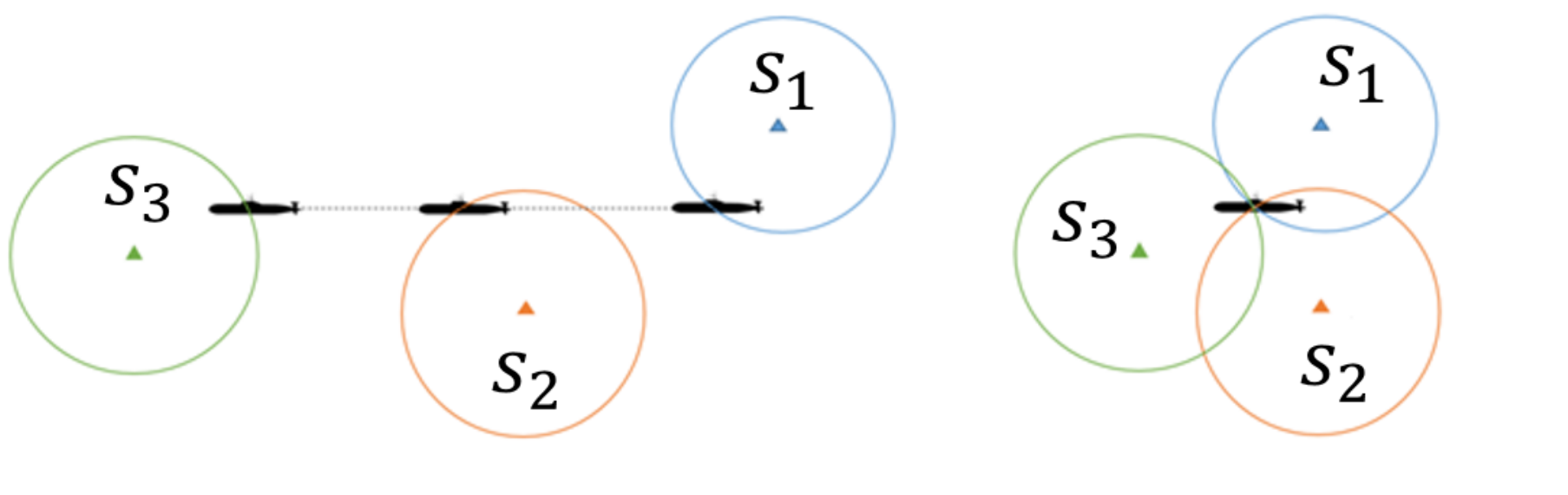}%
\label{fig:range_windows}}
\caption{Ranging windows (a) and sequential trilateration (b).}
\label{fig:range_windows_and_trilat}
\end{figure*}
When ranging technique is applied to satellite signal as in GNSS, the signal itself encodes time information and the parameters to calculate the satellite position. This kind of approach, although theoretically possible using neutrino packets~\cite{Stancil}, is not feasible with our current set-up. Notice that the signal rate for MINER$\nu$A is only 0.1 bits/s with a distance source-detector of 1 km and better conditions in terms of flux with respect to ours. 
Therefore we propose a PNT algorithm based on the concept of Ranging Windows as depicted in Fig.~\ref{fig:range_windows_and_trilat}.
Each neutrino source would fire only during a certain time interval $t_S$ while the others are shut down in order to allow source discrimination.
During the fire interval of each source, $t_S$, the source would fire in pulses (one or more pulses) of duration $t_P$ and spaced by a time interval $t_{BP}$. The time between pulses, $t_{BP}$, is defined so that two pulses cannot overlap in a given coverage area. In fact, the detector will receive the signal coming from the pulse with some delay $\Delta t$, $t_{BP}$ is chosen to be larger than the maximum delay allowed in the coverage region. For instance, if the coverage region is all the Earth the upper bound would be the Earth diameter times the speed of light.  
The length of each pulse is linked to the target accuracy since the exact time of emission would be indistinguishable within $t_P$ intervals.
In this way, if we receive a signal at some time $t$ given by the user clock, we can deduce from which source and which train of pulses it comes from and in this way we can calculate the time delay $\Delta t^{\prime}$ = $\Delta t$ + errors.  
In the case of submarine navigation for an accuracy requirement of 1 km, $t_P$ would be of the order of \num{7e-6}~s. Concerning $t_{BP}$ for a global coverage of all the earth, $t_{BP}\simeq \num{4.2e-2}$~s, while for a coverage area of the order 4000 km, $t_{BP}\simeq \num{1.3e-2}$~s.

The Neutrino PNT User Equipment will be coupled with additional sensors like an IMU. A Sequential Equivalent Trilateration algorithm for the hybridization has been defined in order to take into account the large time between neutrino measurements. The trilateration needed for positioning the submarine is given in different epochs 
(due to the long time between two consecutive neutrino detections), in contrast to instantaneous trilateration (as in GNSS) where the three signals (plus clock) are received at the same instant. The principle is the same, but at each instant the position and its accuracy is updated based on the beam direction and propagated with the inertial sensors until a new neutrino is detected.
Source positions are known in advance and provided by the control segment.

\section{Particle Physics Aspects}
This section discusses in detail the neutrino source identified as the best candidate (\cref{sec:neutrino_sources}), the detector technology and measured process that is paired with this source (\cref{sec:neutrino_detector}), and the major backgrounds that would be seen at the detector (\cref{sec:background}).
Due to the properties of neutrinos already introduced, they have a very low probability of interaction with matter. The low probability of interaction is due to the very short range of the weak force, and the relative strength of the gravitational force being tens of orders of magnitude lower than the weak force. As mentioned in \cref{sec:previous_works}, the interaction via {\cevns} process is used in this work, this process occurs with all flavours of neutrino. 
\subsection{Neutrino Sources}\label{sec:neutrino_sources}
Several neutrino sources exist which have been used to experimentally characterise the nature of the neutrino. Neutrinos are generated in the fusion reactions which power the sun, as the result of particle decays from cosmic ray interactions in the earth's atmosphere, in the supernovae collapse of stars, and at the start of the universe. They are also generated by nuclear fission reactions, and by collisions between particles at accelerators. This section gives some additional detail on neutrinos generated as a result of pion decay at rest.


The generation of neutrinos at accelerators is the result of particle decays, an accelerated proton is incident on a target producing pions, \(\pi\), and other particles. The \(\pi\) and \(\mu\) that are produced then will decay with characteristic decay times \(\tau_{\pi}\) and \(\tau_{\mu}\) respectively 
\begin{equation}
\pi^{+} \rightarrow \mu^{+} + \nu_{\mu},
    \label{eq:pi_decay}
\end{equation} the muon, \(\mu^{+}\), that originates from the pion decay will itself decay 
\begin{equation}
\mu^{+} \rightarrow e^{+} + \nu_{e} + \bar{\nu}_{\mu}.
    \label{eq:mu_decay}
\end{equation} 
Depending on the energy of the incident proton, the \(\pi\) that are produced are either at rest or have some momentum. In the case of a decay at rest the flux that is emitted is isotropic, protons with the required energies can be generated by cyclotrons. For this study the proposed system of compact cyclotrons for the DAE$\delta$ALUS experiment \cite{Conrad2009}, which generates \num{4e22}~\(\nu\)/year/flavour, was used as a reference value. 

When a higher energy pion is produced by a proton beam of greater energy, other techniques can be used. The higher energy pions, and other charged particles can be focussed into a narrow beam by magnetic horns \cite{Ichikawa2012,Adamson_2016}. These beams are not however perfectly collimated, they are cone shaped as the decays of the charged particles to neutrinos have a distribution and diverge over the distance to the detector. The infrastructure required to produce these beams is significant. 

\subsection{Neutrino Detector} \label{sec:neutrino_detector}
\begin{figure}[h]
    \centering
    \includegraphics[width=\linewidth]{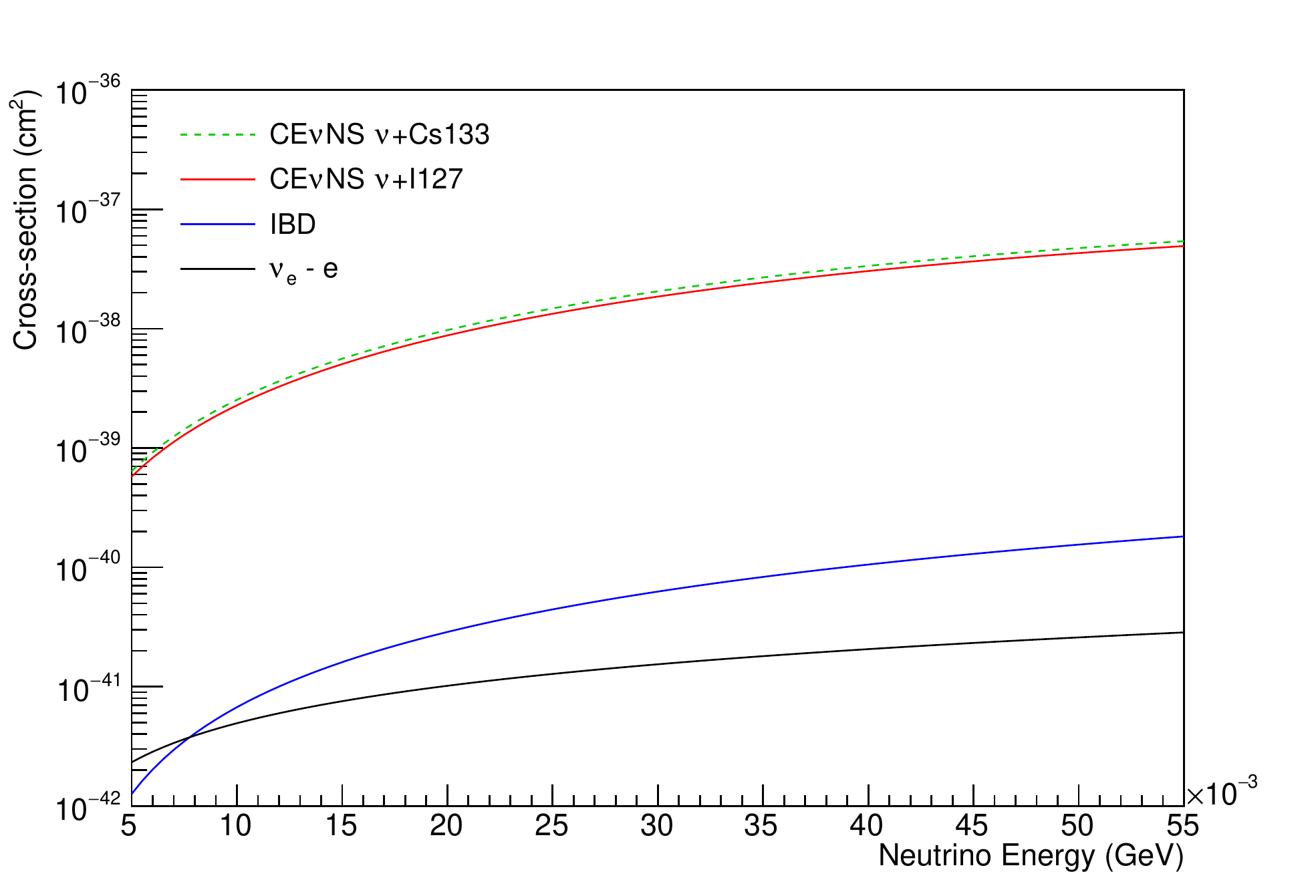}
    \caption{Neutrino cross section as a function of the neutrino energy for \ce{^{133}Cs} (dashed green) and \ce{^{127}I} (red) for the {\cevns} process. Also shown are Inverse Beta Decay (IBD) (blue), and neutrino electron scattering (\(\nu_{e} \,\text{--}\, e \)) (black), cross section predictions from GENIE \cite{Andreopoulos:2015wxa}}
    \label{fig:CEvNS_xsec}
\end{figure}

Until recently, neutrino experiments relied on detectors with target masses of several tonnes to many kilo-tonnes which are required in order to give a reasonable probability of interaction and detect a signal. An experiment by the COHERENT collaboration in 2017 made the first measurement of {\cevns} process with a sensitive CsI mass of \SI{14.57}{\kilo\gram} \cite{Akimov2017}, compared to detectors which have sensitive masses in the range of \SIrange{1}{50e3}{\tonne} \cite{Fukuda:2002uc}. The {\cevns} process has a large cross section at low energies and is proportional to the square of the number of neutrons in the atom \cite{Scholz2019}. By selecting a target \ce{CsI} scintillator crystal, which has large numbers of neutrons in the constituent \ce{Cs} and \ce{I} atoms, the {\cevns} cross section is at least two orders of magnitude larger than other processes at the same energy, and maximises the probability of detection. A comparison of cross sections in the energy region of interest is shown in \cref{fig:CEvNS_xsec} and includes: the {\cevns} process on \ce{^{133}Cs} (dashed green) and \ce{^{127}I} (red) atoms, inverse beta decay (IBD) (blue) which was used in the discovery of the neutrino, and neutrino electron scattering (\(\nu_e \,\text{--}\, e\)) (black). These \ce{CsI} crystals are widely used in the detection of radiation and particle physics experiments. A detector could be constructed from multiple \ce{CsI} crystals, as opposed to the single crystal COHERENT \ce{CsI} experiment, each crystal instrumented by Silicon Photomultipliers/Multi-Pixel Photon Counters (SiPMs/MPPCs), which are more much compact than Photomultiplier Tubes (PMTs). The additional spatial information from the segmented detector would allow for improved background rejection, by identification of of background events which have particular topologies which aren't identified by the muon vet. The SiPM/MPPC technology also has a lower power requirement \num{\sim 2} orders of magnitude compared to PMTs. This leads to a power requirement of order \SI{5}{\kilo\watt} for a detector with \num{1e4} crystals. The shielding that is used to reduce the background, is composed of several layers of different materials which are shown and described in \cref{fig:shielding_layers} and are further discussed in \cref{sec:background}. 

\subsection{Backgrounds} \label{sec:background}
Due to the nature of the interactions that are generated by neutrinos, other particles can have interactions which are difficult to distinguish from those that are generated by neutrinos. In particular neutrons can traverse matter without interaction and can deposit energy in a nucleus generating the same signature nuclear recoil as the {\cevns} reaction. Cosmic ray muons can generate neutrons in interactions that occur with the materials which shield the detector and in the active volume itself, which then mimic the neutrino signal. By placing the detector on a submarine requires consideration of the additional backgrounds from the neutrinos and neutrons that are generated in the reactor core. The backgrounds that are generated by cosmic rays are mitigated by shielding the detectors by locating them underground, typically in mines. In the case of submarines, a variable amount of shielding from water would be present. The background from the reactor neutrons, and to some extent the cosmic rays, can be mitigated by layers of material which surround and shield the sensitive \ce{CsI} crystal. 

\begin{figure}[h]
    \centering
    \includegraphics[width=\linewidth]{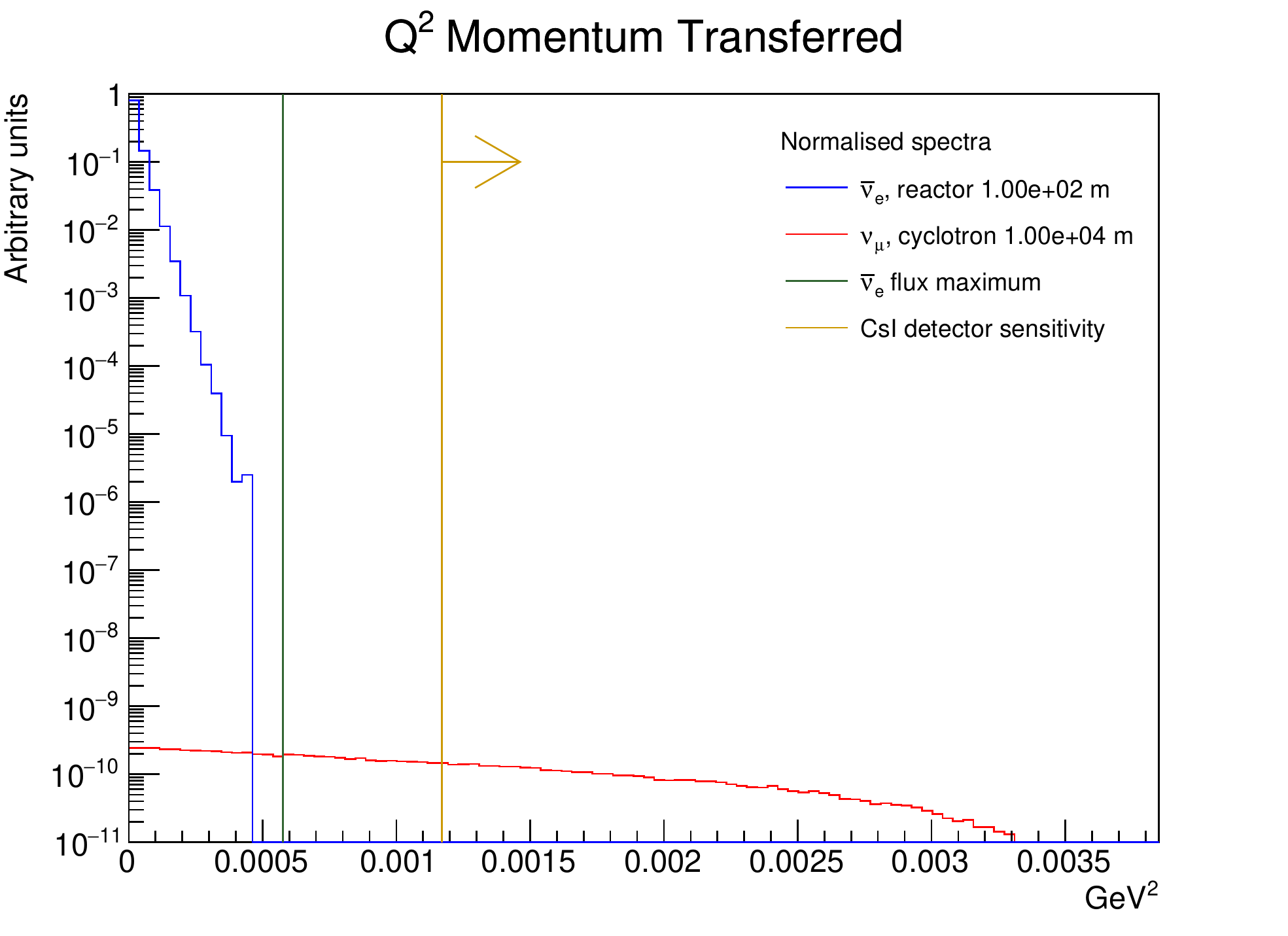}
    \caption{Momentum transferred, \(Q^2\), for the cyclotron pion decay at rest source (red) and the reactor source (blue). Also shown are vertical lines showing the largest \(Q^2\) value for the reactor (green) and the sensitivity of the \ce{CsI} crystal (yellow). The fluxes are scaled for a \SI{200}{\mega\watt} reactor at \SI{100}{\metre} from the detector and \num{4e22}~\(\nu/\mathrm{year}\) at \SI{1e4}{\metre} from the detector for the cyclotron.}
    \label{fig:q2_reactor}
\end{figure}

In order to characterise the background generated by the reactor neutrinos, \(\bar{\nu}_{e}\), a GENIE (Generates Events for Neutrino Interaction Experiments) \cite{Andreopoulos:2009rq,Andreopoulos:2015wxa} simulation is performed of the momentum transfer squared, \(Q^2\), which is the transfer of four momentum from the neutrino to the nucleus, for the reactor neutrino flux and for the flux from the pion decay at rest source. A plot of the distributions, together with the maximum \(Q^2\) for the reactor and the \ce{CsI} detector sensitivity are shown in \cref{fig:q2_reactor}. In \cref{fig:q2_reactor} the neutrino fluxes are scaled for a reactor of \SI{200}{\mega\watt} at \SI{1e2}{\metre} (\(\bar{\nu}_{e}\)) and a flux of \num{4e22}~\(\nu/year\) (\(\nu_{\mu}\))  at \SI{1e4}{\metre} for the cyclotron. The \ce{CsI} crystal is not sensitive below the yellow line so the background from the reactor \(\bar{\nu}_{e}\) is not observed.

\begin{figure}[h]
    \centering
    \includegraphics[width=0.9\linewidth]{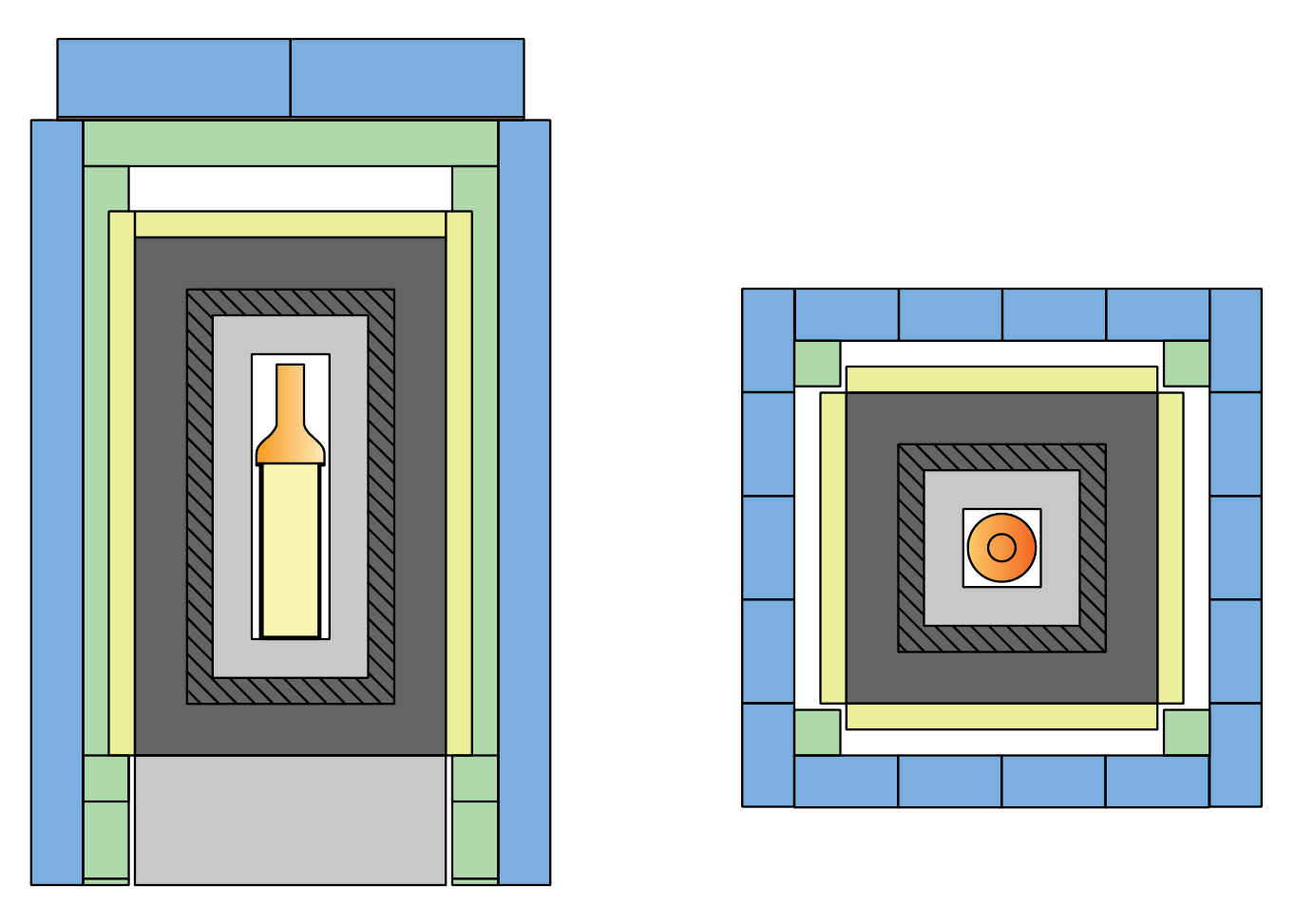}
    \caption{Layers of shielding in the \ce{CsI} COHERENT experiment, side view (left), top view (right). The central \SI{14.57}{\kilo\gram} \ce{CsI} crystal is shown (yellow) in the side view, it is instrumented by a Photomultiplier Tube (orange), the first shielding layer is low background HPDE (grey,  \SI{75}{\milli\metre}), followed by low background lead (hatched dark grey, \SI{50}{\milli\metre}), contemporary lead (grey, \SI{100}{\milli\metre}), scintillator panels (yellow, \SI{50}{\milli\metre}), aluminium extrusion as support structure (green), finally water (blue, \SI{92}{\milli\metre}) from \cite{Scholz2019}}
    \label{fig:shielding_layers}
\end{figure}

In order to simulate the neutron background that is generated by the submarine reactor and those that are induced by cosmic rays in the shielding layers of the detector a GEANT4 \cite{Agostinelli:2002hh,Allison:2016lfl} model of a detector was developed. This model is based on the layers of shielding that were used for the \ce{CsI} experiment in the COHERENT collaboration. Working outwards, the central \ce{CsI} crystal is surrounded by a layer of low background HDPE (\SI{75}{\milli\metre}), low background lead (\SI{50}{\milli\metre}), contemporary lead (\SI{100}{\milli\metre}), scintillator panels (\SI{50}{\milli\metre}), and finally a layer of water (\SI{92}{\milli\metre}) contained in tanks, these are illustrated in \cref{fig:shielding_layers}.

The energy which is deposited by each neutron interaction in each of the layers of the shielding shown in \cref{fig:shielding_layers} is recorded in the simulation. The shielding shifts the energy that is deposited by the neutrons in the \ce{CsI} crystal down below the threshold for detection. The energy spectrum of the reactor neutrons which used as input to the simulation is taken from \cite{Hakenmuller:2019ecb}.

\begin{figure}[h]
    \centering
    \includegraphics[width=\linewidth]{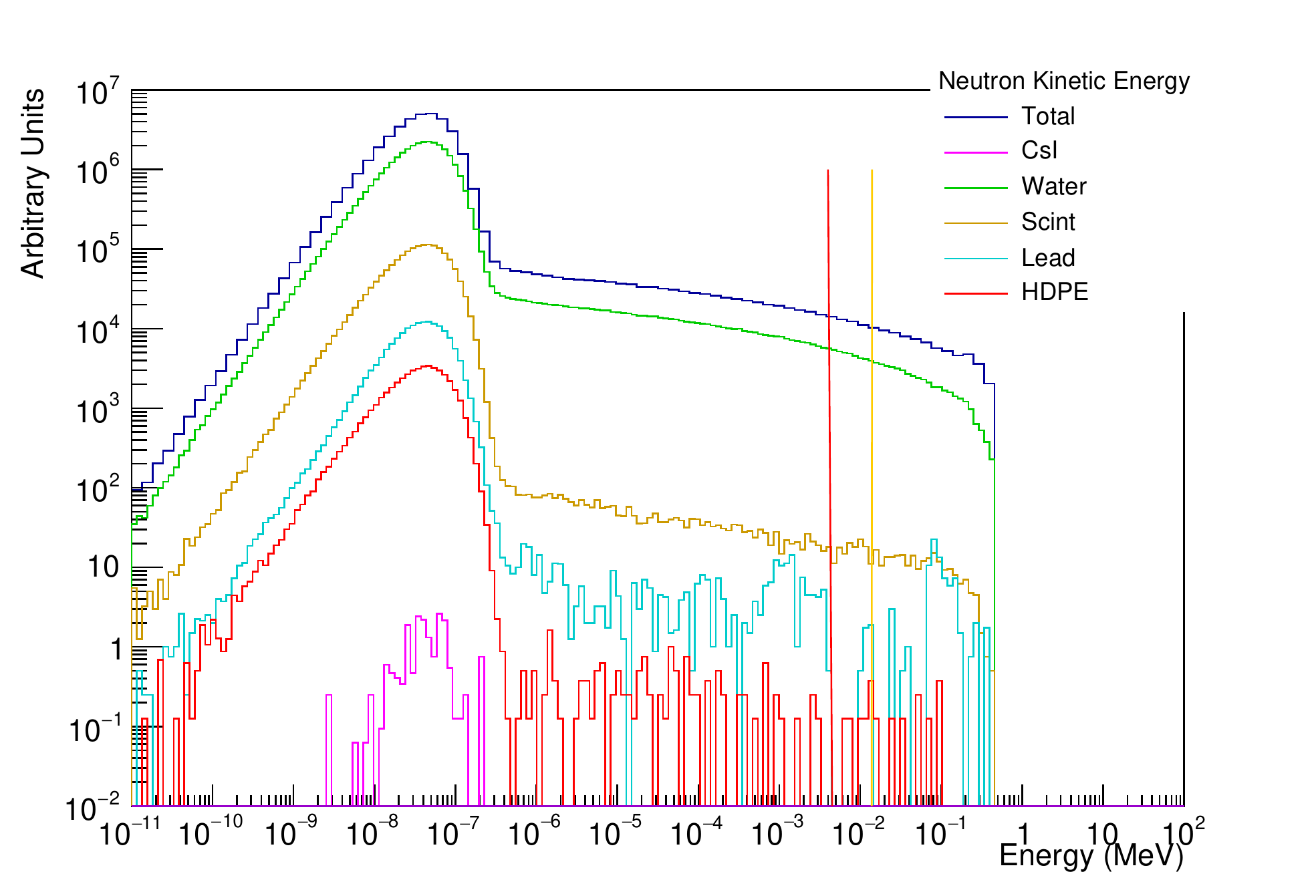}
    \caption{The kinetic energy loss of the neutron in the different regions of the detector. The sensitive \ce{CsI} volume is shown in magenta, the minimum energy to which the detector is sensitive is the vertical red line, and the maximum energy for the {\cevns} process is shown by the vertical yellow line. The other lines show the energy deposited in the insensitive masses of the detector.}
    \label{fig:neutron_kine}
\end{figure}

\begin{figure}[h]
    \centering
    \includegraphics[width=\linewidth]{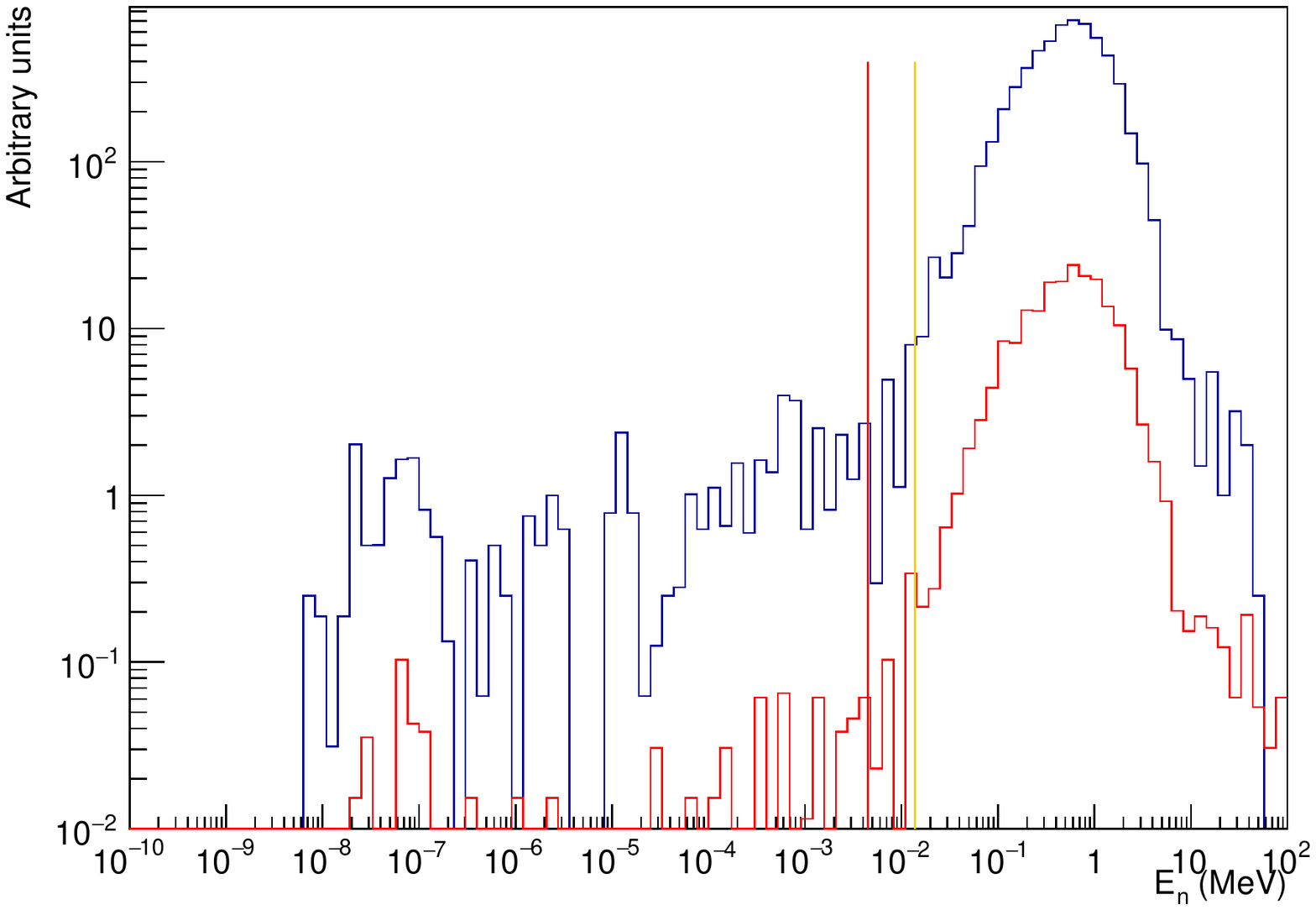}
    \caption{Distributions of the energy deposited by muon induced neutrons in the \ce{CsI} portion of the detector. Two muon energy spectra are used as input to the simulation, one at high energy (red) and a lower energy (blue). The specta are normalised relative to each other based on the number of muons simulated. The red vertical line shows the minimum sensitive energy of the detector, and the yellow vertical line the maximum energy at which the {\cevns} process occurs.}
    \label{fig:muon_induced_neutrons}
\end{figure}

Cosmic rays can induce neutrons in the shielding material which subsequently deposit energy in the \ce{CsI} crystal. The spectrum of cosmic ray muons changes with increasing depth from the surface of the earth, reducing in overall intensity whilst the most probable value shifts to higher energy due to the capture of lower energy muons. Two possible cases are simulated, a shallow case with the spectrum of muons at \num{13}~m.w.e. (metres water equivalent) from \cite{Knei_l_2019}, and a deeper case at \num{\sim 250}~m.w.e. taken from \cite{An:2017jng}. These resulting energy distributions are shown in \cref{fig:muon_induced_neutrons}, these populate the same energy region as the {\cevns} signal, with the higher energy spectrum having a lower intensity in the region of interest. No muon veto was included in the simulation, a veto would allow for a reduction in this class of event and an improvement in the signal identification and would be performed in future work. The segmentation of the detector was not simulated either, this would allow for particle ID of the muons in the sensitive region and veto of those events containing muon like tracks in addition to other events with multiple interactions.  A more detailed future simulation would include this segmentation. The CONUS experiment is located near a commercial fission reactor and aims to measure the {\cevns} cross section from \(\bar{\nu}_{e}\). More detailed simulations performed by the CONUS experiment \cite{Hakenmuller:2019ecb} which has a modest overburden \num{17}~m.w.e shows reductions in the muon background of over an order in magnitude by implementation of a muon veto, a similar reduction in muon background could be expected for the simulations in the current work.

\section{Simulations and Feasability Analysis} \label{sec:sim_feasability}
Two kinds of simulations have been performed for testing the Neutrino PNT concept:
\begin{enumerate}
    \item Particle Physics Simulations for simulating the particle physics processes in the Neutrino generators and detector components in order to estimate the probability of detecting a signal neutrino arriving to the detector. These simulations have taken into account the Neutrino sources physics based on the process Pion decay at rest and the COHERENT detectors processes as proposed in the Baseline Design for the Neutrino PNT System;
    \item Simulations using a PNT tool simulating the Neutrino PNT algorithm and the hybridization with other sensors in order to estimate the PNT performances;
\end{enumerate}

Different configurations representative of different key design parameters of the Neutrino PNT System have been tested:
\begin{itemize}
    \item Neutrino flux generated in each neutrino source in terms of number of neutrinos generated per flavor and per unit time; 
    \item Distance between the sources and the detector, including simulation of the propagation of the flux which decreases as \(1/r^{2}\), \(r\) being the distance source-detector;
    \item Detector weight, which, on the one hand, should be minimized due to operational constraints of the Neutrino PNT applications, but on the other hand, the number of neutrinos detected increases linearly with the detector weight;
    \item Characteristics of the neutrino signal generated, this is, the neutrino pulse length and the spacing between pulses, and the fire windows for the different sources emission, according to the Neutrino PNT concept proposed;
    \item IMU performances typical of submarines;
    \item Accuracy requirement;
    \item Probability of detecting a Neutrino arriving to the detector (provided by the Particle Physics simulations).
\end{itemize}

\subsection{Particle Physics Simulations}
The simulations of the physics processes were broken into two components: one part detailing the time structure of the neutrino pulses, and the other dealing with the event rates expected with the variation of several parameters. Both cases use the {\cevns} cross section to determine the number of events that would be expected on a \ce{CsI} target.

\begin{figure}[h]
    \centering
    \includegraphics[width=\linewidth]{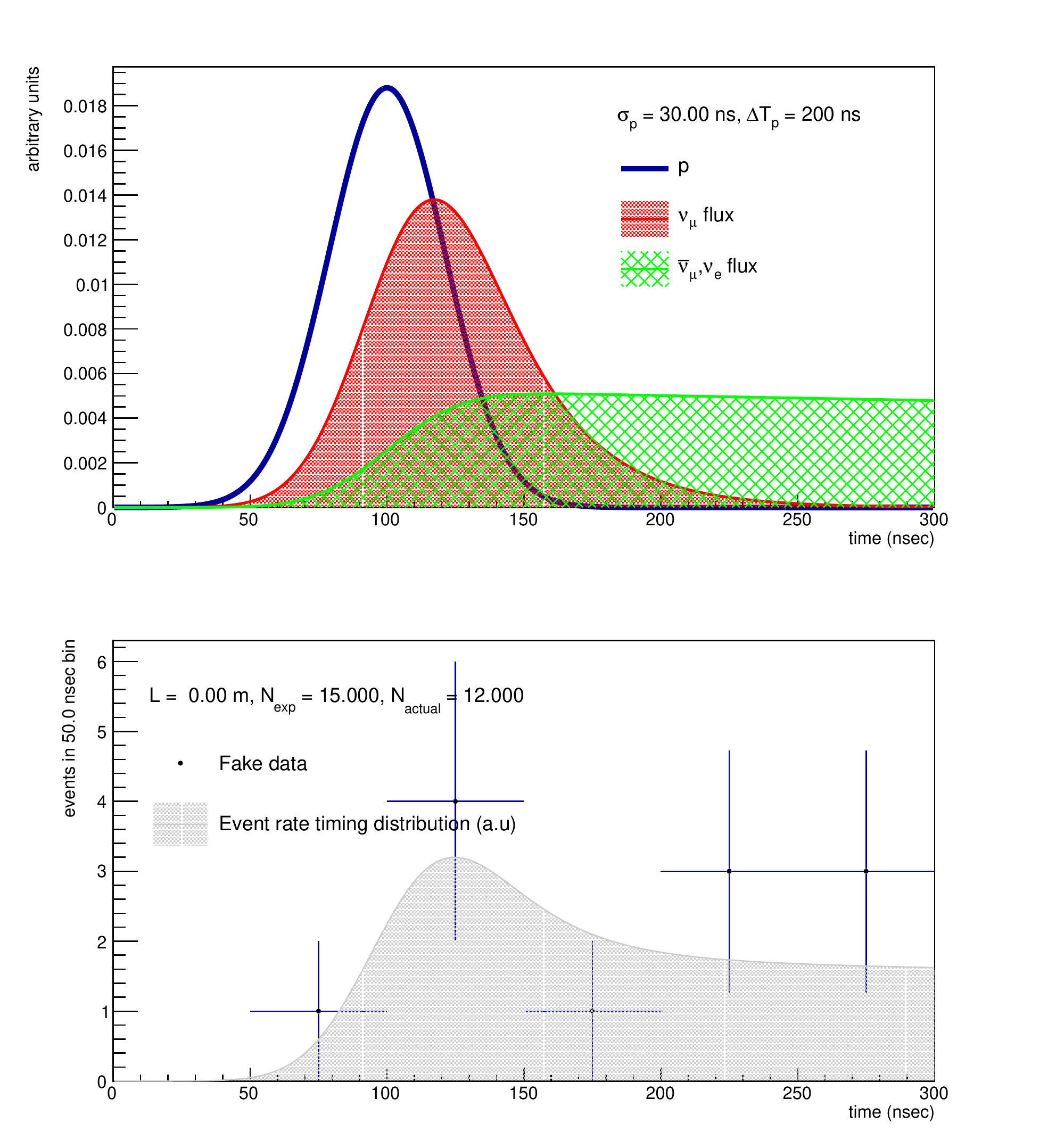}
    \caption{Top panel, generation of the neutrino fluxes from the initial proton pulse (\(\sigma_t = \SI{30}{\nano\second}\)) (blue), the signal \(\nu_{\mu}\) component of the beam (red shaded) results from the pion decay in \cref{eq:pi_decay}, this is followed by the intrinsic \(\nu_{e}\) and \(\bar{\nu}_{\mu}\) (green shaded) from the muon decay \cref{eq:mu_decay}. Bottom panel, simulated events ``fake data" (blue points) randomly drawn from a Poisson distribution according to the total neutrino flux shape multiplied by the {\cevns} cross section (grey shaded) in a particular time interval.}
    \label{fig:event_timing}
\end{figure}

The time structure of the initial proton beam interaction and that of the neutrino fluxes that are created as a result of the decays of \(\pi\) and \(\mu\) described in \cref{eq:pi_decay} and \cref{eq:mu_decay} respectively are simulated. In \cref{fig:event_timing} a simulated gaussian proton beam (blue) impinges on a target generating \(\pi\), the decays of the \(\pi\) and \(\mu\) each have a characteristic half life, the half lives \(\tau_{\pi}\), \(\tau_{\mu}\), are used to populate the time distribution of the fluxes for the \(\nu_{\mu}\) (red),  \(\nu_{e}\) and \(\bar{\nu}_{\mu}\) (green). The {\cevns} cross section is equally sensitive to all flavours of neutrinos.  The \(\nu_{e}\) and \(\bar{\nu}_{\mu}\) are an intrinsic beam background due to their width in time being much broader than the \(\nu_{\mu}\) from the \(\pi\) decay \(\tau_{\pi}\SI{\sim 26}{\nano\second}\) \cite{Zyla:2020zbs}. For the purposes of the PNT system the longer muon decay time \(\tau_{\mu}\SI{\sim 2.2}{\micro\second}\) \cite{Zyla:2020zbs} can mean that \(\nu_{e}\) and \(\bar{\nu}_{\mu}\) pulses overlap leading to uncertainty in timing and hence positioning in addition to an spatial uncertainty at least two orders of magnitude larger. The timing of the source pulses, repetition and width, can be adjusted so that the background component of the pulses can decay before the next signal pulse, in order to meet the criteria described in \cref{sec:PNTAlg}.
The flux distribution generated in the previous step is multiplied with the {\cevns} cross section for \ce{CsI} which can be seen in bottom panel of \cref{fig:event_timing}. The resulting distribution is then assigned a total number of expected events, and the number of simulated expected events is drawn from a Poisson distribution for a particular time interval of the distribution when events are observed seen as points in the bottom panel of \cref{fig:event_timing}.

Event rates are calculated taking into consideration a variety of parameters which are detailed in \cref{sec:sim_feasability}; 2D distributions of the event rates were produced as a function of the distance and the active mass of the detector. The calculation of an event rate, \(N_f\), through a detector requires solving a complex integral but for the studies performed here a simplified model is used:
\begin{equation}
    N_f = \int \frac{d^3 \phi}{dE\,dS} \sigma(E) \frac{N_A}{A} dE\,dm,
\end{equation}
where \(d^3 \phi/dE\,dS\) is the neutrino energy flux per energy bin, \(dE\), per unit area \(dS\), \(\sigma(E)\) is the energy dependent {\cevns} cross section, \(N_A\) is Avogadro's constant, \(A\) the mass number and \(m\) the mass. The integral is evaluated computationally
\begin{equation}
    N_f = \frac{N_A}{A} M_{fv} E_{bw} \sum^{n_\mathrm{bins}}_{i=0} F_i \sigma_i,
\end{equation}
where \(M_{fv}\) is the mass of the detector, \(E_{bw}\) is the width of the energy bin, \(F_i\) is the flux and \(\sigma_i\) the cross section in bin \(i\). The pion decay at rest spectrum is used as the flux shape. To account for the exposure of the detector to the neutrino flux a scaling \(\phi_t\), which is based on the pulse structure of the beam (repetition time and pulse width), is applied. A scaling for distance, \(1/r^{2}\) is applied with the simplification that the flux is fully collimated when it impinges on the detector, and the detector is spherical with the corresponding area it would present to the flux. 
The probability of a neutrino interacting which arrives at a detector was also evaluated with a value of, \SI{9.43e-14}{\per\kilo\gram}, and is used as an input to the PNT simulations which are detailed in \cref{sec:PNT_sim}.

\begin{figure}[h]
    \centering
    \includegraphics[width=\linewidth]{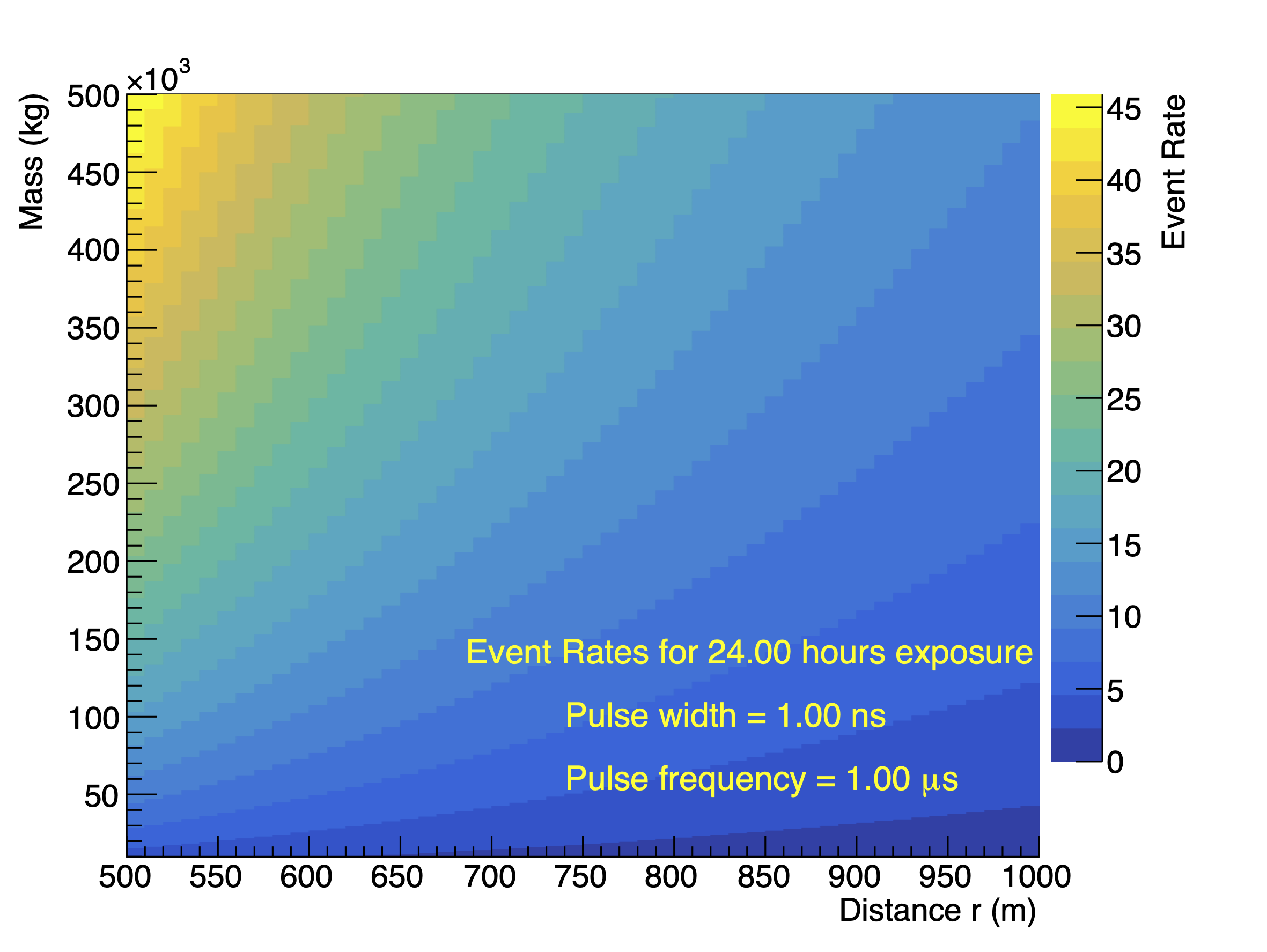}
    \caption{The 2D distribution of events expected for \num{24}~hours with a pulse width of \SI{1}{\nano\second} and a pulse frequency of \SI{1}{\micro\second}, as a function of the distance from the source, and the mass of the sensitive part of the detector. This plot shows a scan of the mass range \SIrange{5e2}{5e3}{\kilo\gram}, and distance range \SIrange{5e2}{1e3}{\metre}, giving a maximum number of events expected of \num{\sim 45}. }
    \label{fig:event_rates}
\end{figure}

\begin{figure}[h]
    \centering
    \includegraphics[width=\linewidth]{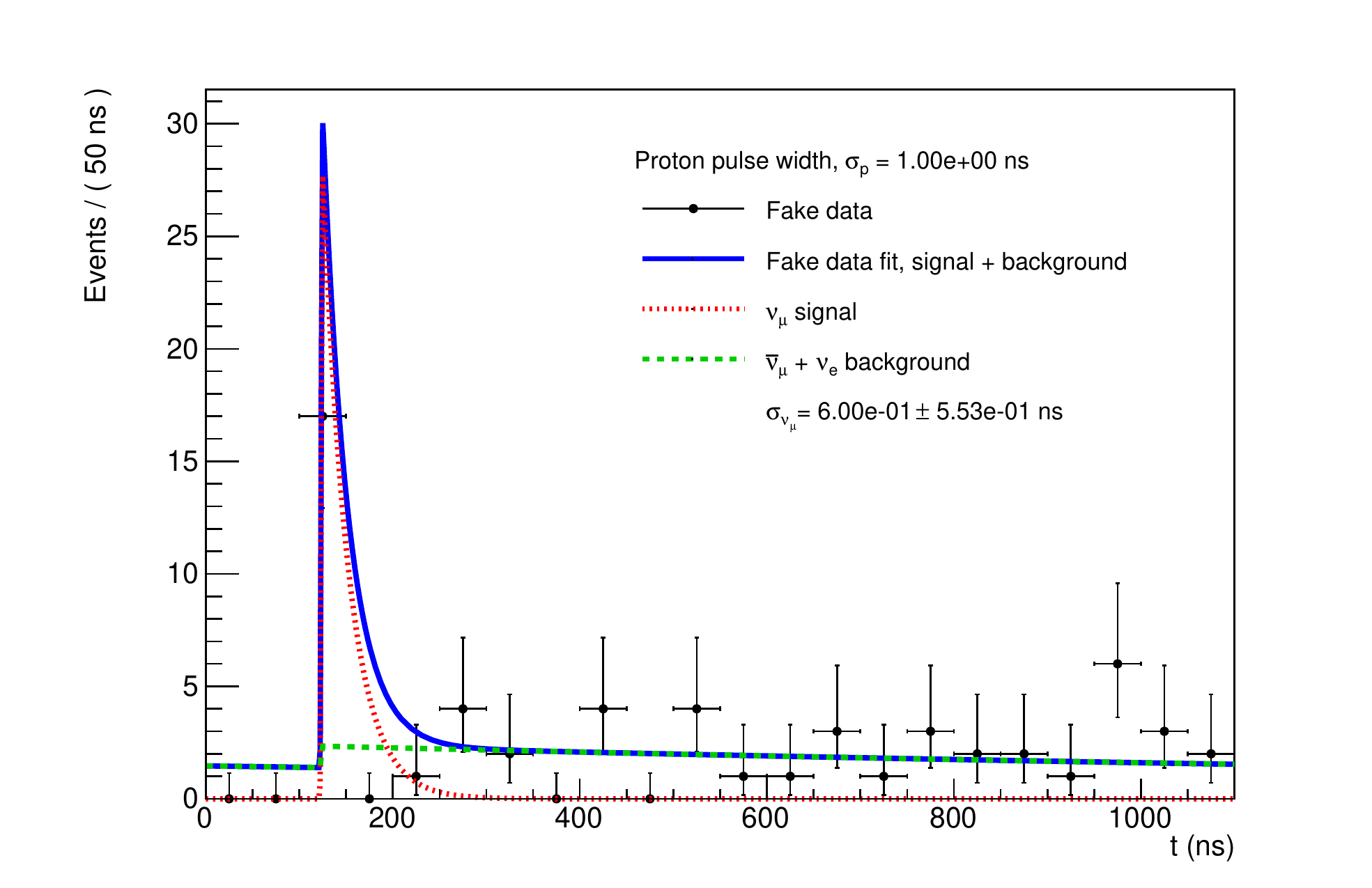}
    \caption{Fit to simulate the experimentally measured width of pulse, giving a timing uncertainty, that could be measured using simulated data points from the cyclotron pulses that are accumulated within a time window. The knowledge of the decay times \(\tau_{\pi}\), and \(\tau_{\mu}\), and the correlation of background and signal region peaks are used to constrain the parameters to fit the decays. The number of simulated events that are expected, along with the pulse width and repetition from \cref{fig:event_rates} are used. The width of the pulse that is found is \SI[separate-uncertainty=true]{0.600(553)}{\nano\second}}
    \label{fig:neutrino_fit}
\end{figure}

A fit is carried out to determine the width of the pulse that would be detected as a result of the signal, as this forms a major part of the spatial resolution which can be achieved by the PNT algorithm. The measured data points are accumulated within a certain time window to allow for a distribution to be accumulated which can be fit. This fit models the flux in the signal and background regions as the convolution of a gaussian and exponential decay, and uses knowledge of the pion and muon half lives \(\tau_{\pi}\), \(\tau_{\mu}\) respectively, and the time correlation of the background and signal peaks to constrain the fit parameters. The timing uncertainty of the neutrino pulse is taken as the width of the signal gaussian which is measured as the result of a fit to simulated data points, an example of this fit is shown in \cref{fig:neutrino_fit}. 

\subsection{PNT Simulations} \label{sec:PNT_sim}
The PNT simulations were configured according to the parameters shown in the following table.
\begin{center}
\begin{tabular}{ m{5.5cm}|m{2.5cm} } 
 \toprule
 Configuration Parameter & Value \\ 
 \midrule
 Typical distance Neutrino source - detector & 1000 km \\ 
 \midrule
 Probability of detection of a signal neutrino arriving to the detector & \SI{9.43e-14}{\per\kilo\gram} of detector \\ 
\midrule
 Ranging window parameter time of pulse, according to \SI{1}{\kilo\metre} accuracy requirement  & \SI{7e-6}{\second} \\
\midrule
 Neutrino detector weight & 500 tons \\
\midrule
 Performances of Inertial Measurement Unit (IMU) typical of Submarines & Position (TDRMS): 1NM/24h \\
 \bottomrule
\end{tabular}
\end{center}

The configuration parameters have been selected in such a way to be representative of a realistic Neutrino PNT system. The typical distance neutrino source - detector has been considered in such a way to need the order of 10 neutrino sources to cover the Mediterranean sea. In the design of a Neutrino PNT system, a compromise between cost and performances would have to be established. The probability of detection of a signal neutrino arriving to the detector has been configured according to the particle physics simulations' results. The ranging windows parameters have been configured according to 1km accuracy requirement with a service coverage in an area of the order thousands of kilometers. IMU performance has been configured according to the typical performances of submarine IMU. The neutrino detector weight has been configured taking into account that typical submarine masses could be of the order of 10.000 tons   \\
In the simulations it has been observed that an IMU-standalone solution is degraded until the target accuracy requirement is not met while the hybrid IMU + Neutrinos PNT solution achieves a stationary state. \\
\begin{figure}[!t]
\centering
\includegraphics[width=\linewidth]{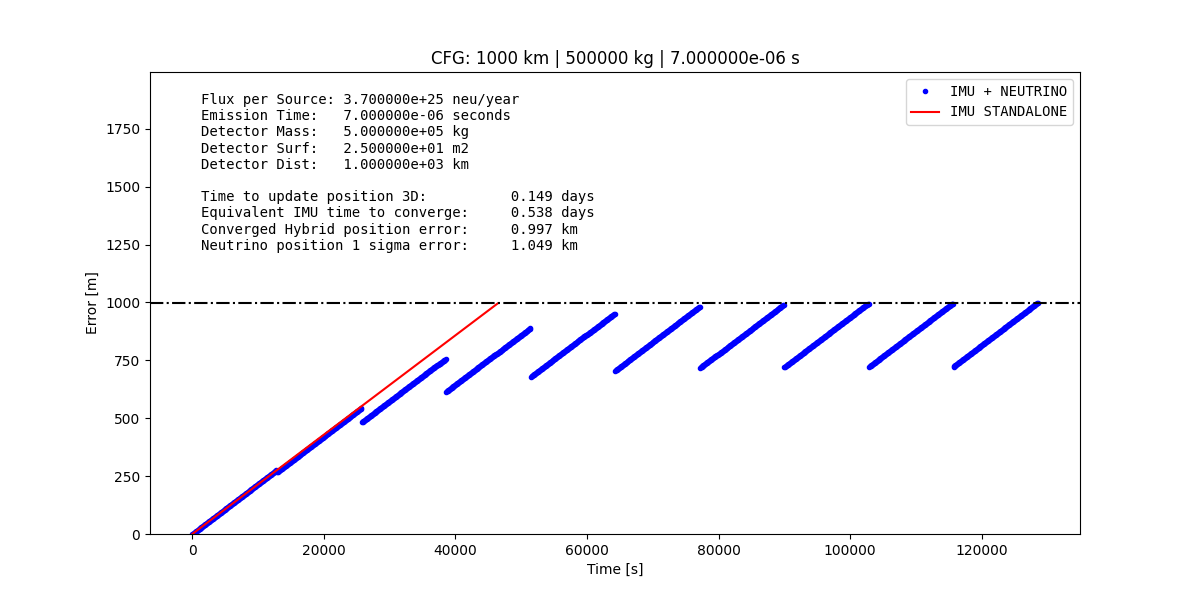}
\caption{PNT Simulation Results}
\label{fig:1000km_500000kg_0000s}
\end{figure}
The results of the PNT simulations show that for complying with 1 km accuracy requirement, a generated neutrino flux of the order \num{3.8e25} $\nu$ / year is needed, as shown in \cref{fig:1000km_500000kg_0000s}. 

\subsection{Feasability Analysis}
The results of the PNT experimentation show that the feasibility of the Neutrino PNT concept is not that far with current technology. In particular, for submarine applications, a Neutrino PNT system could provide an added value if a neutrino flux of 3 orders of magnitude larger than the one considered in a reference experiment such as DAE$\delta$ALUS \cite{Conrad2009} is achieved. The European Spallation Source neutrino Super Beam (ESSnuSB) \cite{ESSnuSB} aims to use the linear accelerator at the European Spallation Source with upgrades to operate at a power of 5 MW and at an energy of 4 GeV producing \num{5.08e23} v/year, one order of magnitude larger than DAE$\delta$DALUS in the same energy range. Taking ESSnuSB as a reference experiment, the simulation results show that a larger neutrino flux of 2 orders of magnitude than ESSnuSB would be needed to comply with the accuracy target. It is important to remark that we have been conservative with respect to the generated Neutrino flux, which is the key parameter to improve the performances of the Neutrino PNT system. We have considered the number provided in the design of DAE$\delta$ALUS and ESSnuSB experiments but this does not mean that these numbers correspond to the current cutting-edge technology in particle accelerators. In addition the generated neutrino flux could be improved in the future with new particle accelerators design within certain physical and operational limits.
Due to these arguments, we believe that the feasibility of a Neutrino PNT concept could be at the limit as supported by the results of this first-of a kind study in Europe. The results of the experimentation show the difficulties to build a Neutrino PNT system but are not that far to make this innovative idea feasible. We believe that in the next decade the Neutrino PNT system may be feasible if:\\
\begin{itemize}
    \item Neutrino detector technology based on the Coherent elastic neutrino-nucleus scattering is improved for enhancing the neutrino detector rate.
    \item And considering cutting edge technology for the Neutrino sources, possibly with a need of evolution in the next years in order to enhance the intensity of the generated neutrino flux. 
\end{itemize}

\section{Conclusion}
In this paper we analysed the feasibility of a PNT system based on neutrinos for those special cases where conventional navigation systems have serious limitations or do not operate in the first place. One such scenario, studied in detail in this project, is that of submarine navigation. Our proposed design is based on a set of fixed isotropic neutrino sources and detectors based on Coherent Elastic neutrino-nucleus scattering. The use of detectors taking advantage of this effect is crucial to reduce the total mass of the on-board neutrino detectors. Without it, the weight and mass of the detectors alone would render the entire idea completely unfeasible. Together with the hardware required for the neutrino based PNT system, this paper proposes a localisation algorithm based on ranging windows concept. Lastly, based on a simulation campaign, it was demonstrated that the proposed navigation system offers several clear benefits over inertial navigation alone, one of which is longer underwater operations. One of the big concerns with neutrino navigation is the effect of neutrino backgrounds on the navigation system, especially taking into account that for nuclear submarines a strong background is expected. Our simulations show that backgrounds, including those created by the nuclear propulsion systems in the submarine, can be distinguished from the real signal and do not pose a threat to the feasibility of the proposed system.

Although the technology required is still not mature enough for the development of a neutrino based PNT, we expect that in the time frame of 10 to 20 years, technological progress will make it possible to develop a proof-of-concept system.

\appendix


 \bibliographystyle{elsarticle-num} 
 \bibliography{positrino_bib}





\end{document}